\documentstyle[aps,twocolumn]{revtex}
\makeatletter
\input{epsfig.sty}
\makeatother

\begin{document}
\draft
\title{Efficient Raman Sideband Generation in a Coherent Atomic Medium}
\author{A.F. Huss, N. Peer, R. Lammegger, E.A. Korsunsky, and L. Windholz}
\address{Institut f\"{u}r Experimentalphysik, Technische Universit\"{a}t Graz, A-8010
Graz, Austria}
\date{\today{}}
\maketitle

\begin{abstract}
We demonstrate the efficient generation of Raman sidebands in a medium
coherently prepared in a dark state by continuous-wave low-intensity laser
radiation. Our experiment is performed in sodium vapor excited in $\Lambda $
configuration on the D$_{1}$ line by two laser fields of resonant
frequencies $\omega _{1}$ and $\omega _{2}$, and probed by a third field $%
\omega _{3}$. First-order sidebands for frequencies $\omega _{1}$, $\omega
_{2}$ and up to the third-order sidebands for frequency $\omega _{3}$ are
observed. The generation starts at a power as low as 10 microwatt for each
input field. Dependencies of the intensities of both input and generated
waves on the frequency difference ($\omega _{1}-\omega _{2}$), on the
frequency $\omega _{3}$ and on the optical density are investigated.
\end{abstract}


\section{Introduction}

Nonlinear optics assisted by electromagnetically induced transparency (EIT)
\cite{harr97} has attracted a great deal of attention in recent years. The
effect of EIT is due to creation, via interference of the excitation paths,
of a coherent superposition of quantum states which does not participate in
the atom-field interaction (''dark'' state) \cite{cpt}. The preparation of
atoms in this superposition gives rise to a strongly reduced absorption and
refraction of the medium, and at the same time, it may lead to enhancement
of the nonlinear optical susceptibility \cite{harr90}. Therefore, nonlinear
optical processes are very efficient in such a coherently prepared medium
(which is called sometimes as ''phaseonium'').

There are several directions in the research of EIT-assisted nonlinear
optics which are actively developed at present. One is an efficient
nonlinear frequency conversion and generation of coherent electromagnetic
radiation unattainable by conventional methods. For example, up-conversion
to VUV wavelengths outside the transparency window of most birefringent
nonlinear crystals has been experimentally demonstrated with unity
photon-conversion efficiency \cite{gener,mer99}; and a high-efficient scheme
for generation of c.w. terahertz radiation by use of EIT has been proposed
\cite{thz}. Another interesting application of the coherent medium concept
is a laser frequency modulation by parametric Raman sideband generation,
with a total bandwidth extending over the infrared, visible, and ultraviolet
spectral regions, and with a possiblity of subfemtosecond pulse compression
\cite{sok,hak}. The third very promising direction is based on the fact that
the intensity ''threshold'' for EIT is given by the decay rate of the dark
state, which can be made extremely small if the dark state is a
superposition of atomic ground state sublevels. For example, a rate below 40
Hz has been observed in a cell with buffer gas \cite{wyn97}, and of order of
1 Hz in a cell with antirelaxation coating \cite{m-o1}. Then, the necessary
intensity corresponds to only a few photons per atomic cross section \cite
{harr98,harr99,luk99}. Therefore, one may expect a highly nonlinear response
of phaseonium to applied e.m. fields at very low intensity levels, even at a
few photon level. The potential of phaseonium has been demonstrated in a
recent series of exciting experiments where the group velocity for a light
pulse of the order of a few meters per second have been measured \cite
{m-o1,hau99,kash99}. Such a slow light propagation velocity is an indication
of huge optical nonlinearities necessary for strong interactions between
very weak optical fields. Recent theoretical work shows that this regime can
be used for photon switching \cite{harr98}, for quantum noise correlations
\cite{luk99}, for generation of nonclassical states of the e.m. field and
atomic ensembles, including entangled states \cite{agar93,luk00a,luk00b},
and for quantum information processing \cite{luk00a,luk00b,flei00}.

In the present paper we report on the experimental observation of an
EIT-assisted nonlinear optical process which combines the latter two
directions: generation of a broad spectrum of Raman sidebands in a medium
with small dark state decay rate, hence having a high efficiency even at low
input light intensities. The process occurs in an atomic medium interacting
with laser radiation in a $\Lambda $ scheme (Fig. 1).
\begin{figure}[htb]
\begin{center}
\epsfig{figure=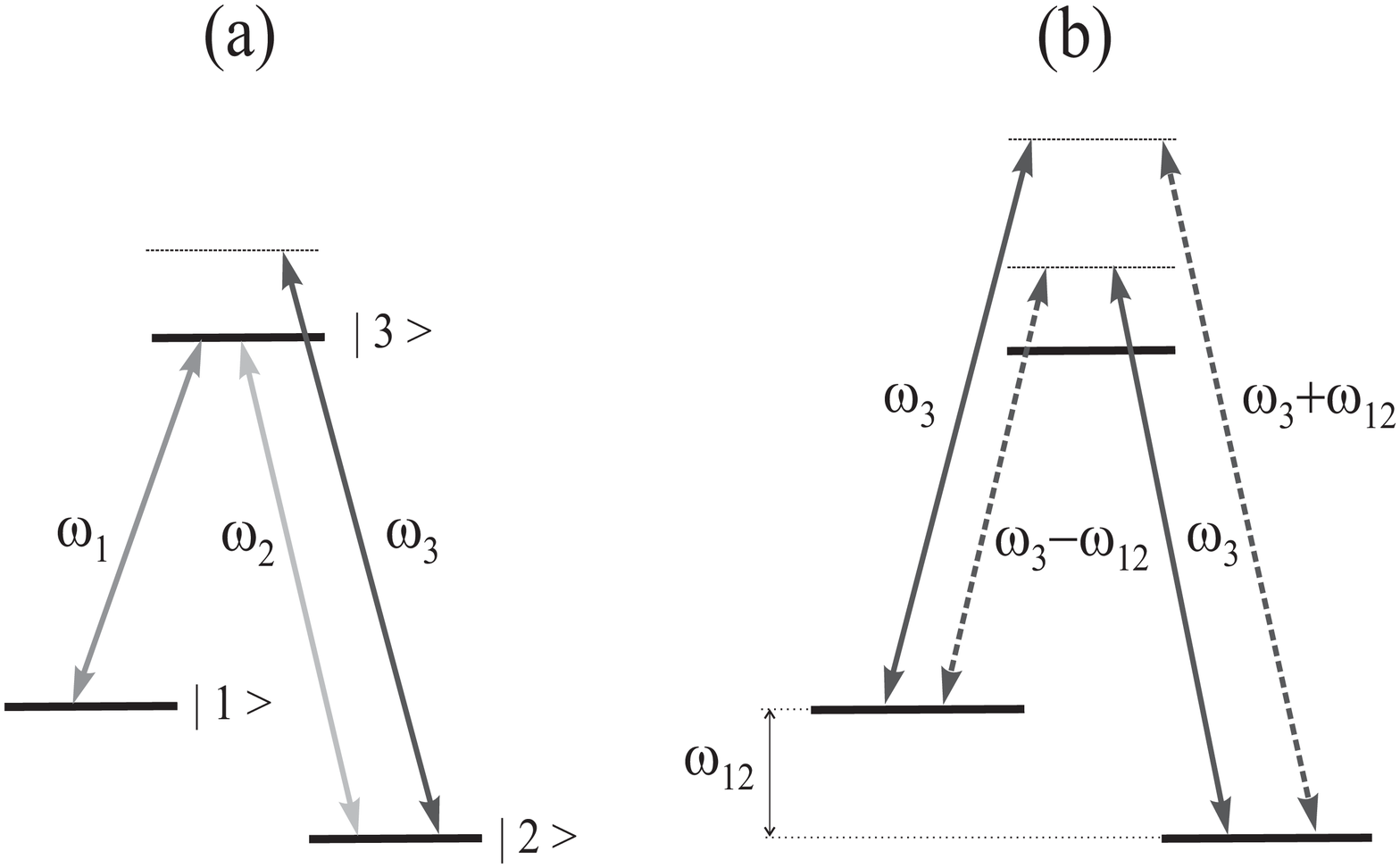,width=7cm}
\end{center}
\caption{\small{$\Lambda $ system in atoms used in our experiment. (a) The $%
\Lambda $-medium is prepared in the dark state by the resonant
pair of fields $\omega _{1}$, $\omega _{2}$ and is probed by the field $\omega _{3}$%
. (b) Schematic demonstration of the generation of Raman sidebands of the $%
\omega _{3}$ field.}} \label{fig1}
\end{figure}
In this system, a pair
of laser frequencies $\omega _{1}$ and $\omega _{2}$ resonantly excites the $%
\Lambda $ system (Fig. 1(a)), which leads to creation of a dark
superposition of both ground states $\left| 1\right\rangle $ and $\left|
2\right\rangle $ and to the preparation of the atoms in this state via
optical pumping. Atoms in the dark superposition act as a local oscillator
at frequency $\omega _{12}$ of the Raman transition. When a third frequency $%
\omega _{3}$ is applied to the system, it will beat against the local
oscillator to produce sum and difference frequencies $\omega _{3}\pm n\cdot
\omega _{12}$, i.e., {\it Raman sideband frequencies} (Fig. 1(b)). $\omega
_{3}$ may be either of the resonant  frequencies ($\omega _{1}$ or $\omega
_{2}$) applied off-resonance to the conjugated transition ($\left|
2\right\rangle -\left| 3\right\rangle $ or $\left| 1\right\rangle -\left|
3\right\rangle $, respectively). Otherwise, it might be derived from an
independent laser and tuned either on resonance or off resonance with one of
the transitions to some additional upper state (in this case, the
interaction scheme is called a double $\Lambda $ scheme). In our experiment,
we have realized these possibilities with all pump waves $\omega _{1}$, $%
\omega _{2}$ and $\omega _{3}$ propagating collinearly, and we have observed
the manifold of the Raman sidebands for all three applied frequencies.

One should note that the use of a coherently prepared medium for the
efficient generation of Raman sidebands has also been theoretically proposed
in Ref. \cite{sok}, and very recently experimentally realized \cite{hak}. In
this work, however, the medium (molecular hydrogen and deuterium) is
prepared in the dark state by far-off-resonance pulsed radiation, which
requires very high intensities. At the same time, the total electromagnetic
energy dissipation is very small by virtue of the large detuning from any
(molecular) state. Therefore, a very broad spectrum of Raman sidebands can
be generated. In our scheme, in opposite, the coherent medium is prepared by
resonant low-intensity c.w. radiation. Since in this case the preparation
relies on the dissipative process of optical pumping \cite{kor97}, a part of
the energy is lost. Nevertheless, the generation is still quite efficient
even at very low pump powers. Besides our present work, the generation of a
single Raman sideband (Stokes field for frequency $\omega _{1}$) in coherent
medium prepared by low-intensity c.w. field has been observed in Ref. \cite
{kash99} for collinear geometry, and several Raman sidebands have been seen
in the related experiment of Ref. \cite{zib99} on parametric
self-oscillation with the counterpropagating driving waves. The aim of the
present experiment is to observe the generation of a broad spectrum of Raman
sidebans in collinear geometry and to study the dependence of each of them
on input laser parameters as well as on the optical density of the medium.

\section{Experimental setup}

Our $\Lambda $ system is generated by the excitation of sodium atoms in a
vapor cell. The lower states are the two hyperfine levels $F=1$ and $F=2$,
spaced by $\omega _{12}=1771.626$ MHz, of the ground state $3^{2}S_{1/2}$,
while the upper state is the hyperfine level $F^{\prime }=2$ of the excited
state $3^{2}P_{1/2}$. The vapor cell contains condensed Na and is
additionally filled with He as puffer gas at a pressure of 6 torr (at room
temperature). It is of cubic form with the length of 10.3 mm and is made of
sapphire to avoid darkening of the cell windows. To compensate stray
magnetic fields the Na cell is placed inside an arrangement of three
mutually orthogonal Helmholtz coils. For the same reason the heating wires
of the cell oven are made of non-magnetic material (Ta) and winded
bifilarly. The cell temperature is electronically controlled and stabilized
to an absolute accuracy of better than 1 ${{}^{\circ }}$C. We have performed
experiments for temperatures ranging from 100${{}^{\circ }}$C to 230${%
{}^{\circ }}$C which corresponds to the saturated Na vapor density from 3$%
\cdot $10$^{9}\,$cm$^{-3}$ to 2$\cdot $10$^{13}\,$cm$^{-3}$. The optical
density $\tau $ has been determined via the absorption of a very weak laser
beam tuned on resonance with $3^{2}S_{1/2},F=2-3^{2}P_{1/2},F^{\prime }=2$
transition: at $\tau =1$, a power of this beam falls by the factor $1/e$.

\begin{figure}[htb]
\begin{center}
\epsfig{figure=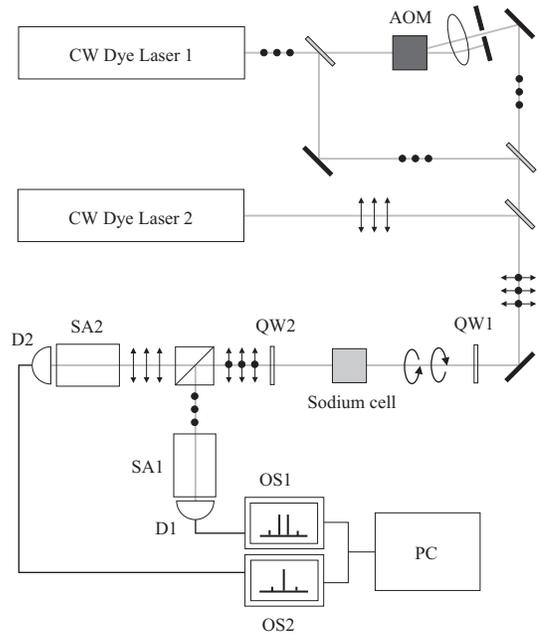,width=7cm}
\end{center}
\caption{\small{Scheme of the experimental setup. See description
in the text.}} \label{fig2}
\end{figure}

The three-frequency radiation is produced by two independent Ar$^{+}$%
-laser-pumped dye lasers with a linewidth of about 1 MHz (Fig. 2). The first
laser system is used to produce EIT in the medium by resonant excitation of
the $\Lambda $-system $F=1-F=2-F^{\prime }=2$. The frequency of this first
laser is stabilized to the $F=2-F^{\prime }=2$ hyperfine transition of the D$%
_{1}$ line (frequency $\omega _{1}$) with a frequency accuracy of 2-3 MHz by
use of saturation spectroscopy on an external temperature-stabilized Na
cell. A part of the beam from this laser, which is linearly vertical
polarized, is split off and led through an acousto-optical modulator (AOM)
driven by a precise tunable RF-generator at 1700-1800 MHz (resolution of
less than 1Hz). The first order sideband produced by the AOM is used as
radiation with the frequency $\omega _{2}$. When an AOM modulation frequency
of 1771,626 MHz is used, the first order sideband is exactly resonant to the
$F=1-F^{\prime }=2$ transition, which corresponds to the Raman resonance
necessary for establishment of EIT. A second cw-dye laser system identical
to the first one provides the laser beam of frequency $\omega _{3}$ with
horizontal linear polarization. The frequency $\omega _{3}$ can be tuned
through the (Doppler broadened with an FWHM of the order of 1 GHz) resonance
$3^{2}S_{1/2},F=2-3^{2}P_{1/2}$ in well defined steps over a range of 32 GHz.

The three laser beams are collinearly overlapped and circularly polarized by
a quarter wave plate (QW1). Thus the frequencies $\omega _{1}$ and $\omega
_{2}$ have in the cell the same circular polarization ($\sigma ^{+}$). This
polarization configuration was found in our preliminary experiments to
produce the best EIT conditions at small intensities. The third frequency $%
\omega _{3}$ is circularly polarized in the opposite direction ($\sigma ^{-}$%
). The combined light beam has a Gaussian transversal profile with a waist
of 0.8 mm, and is almost parallel inside the cell. The input power can be
adjusted with a neutral-density filter, and measured before the cell by a
photo diode (not shown in Fig. 3). After passing the Na cell, the light is
again linear polarized by a second quarter-wave plate (QW2) and the beams of
opposite polarizations are separated by use of a polarizing beamsplitter
cube (PBS). Each of the beams passes an optical spectrum analyzer (scanning
Fabry-Perot interferometer with a free spectral range of 2 GHz, SA1 and SA2)
and is detected on separate photo diodes (D1 and D2) connected to a storage
oscilloscope (OS1 or OS2). The oscilloscopes are read out by a data
acquisition system on a computer (PC). This setup allows us to observe each
frequency component, both input and generated ones. The waves $\omega _{1}$
and $\omega _{2}$ and their generated Raman sidebands (having $\sigma ^{+}$
polarization in the cell) are detected by the system (SA1, D1, OS1), while $%
\omega _{3}$ and its Raman sidebands (having $\sigma ^{-}$ polarization in
the cell) are detected by the second system (SA2, D2, OS2). We should note
that the Raman sidebands have been observed in a setup with all three input
frequencies having the same $\sigma ^{+}$ polarization, too. However, the
efficiency was lower because in this case part of the atomic population is
optically pumped into the state $F=2,m_{F}=+2$ which is not excited by the
applied fields. Moreover, the oscilloscope picture was overcrowded and its
analysis was much more complicated.

\section{Results}

The first step of our experiments was the measurement of the dark state
relaxation rate $\Gamma $. For this purpose, the frequency $\omega _{3}$ was
blocked, and the transmission of frequencies $\omega _{1}$ and $\omega _{2}$
has been measured as a function of the AOM modulation frequency. The result
is a typical EIT transmission peak at 1771,6 MHz, with a halfwidth given by $%
\delta _{EIT}=\Gamma +C\cdot I$, where $I$ is the total intensity of the
frequencies $\omega _{1}$ and $\omega _{2}$, and $C$ is some constant \cite
{cpt}. Thus, the axis offset value of the linear fit to the measured
dependence $\delta _{EIT}(I)$ can be used as an upper limit for $\Gamma $.
In this way, we obtained that the value of $\Gamma $ in our setup is below 3
kHz, which is determined by an AOM frequency jitter and a transit time
broadening (due to finite diffusion time of atoms through the beam).

\begin{figure}[htb]
\begin{center}
\epsfig{figure=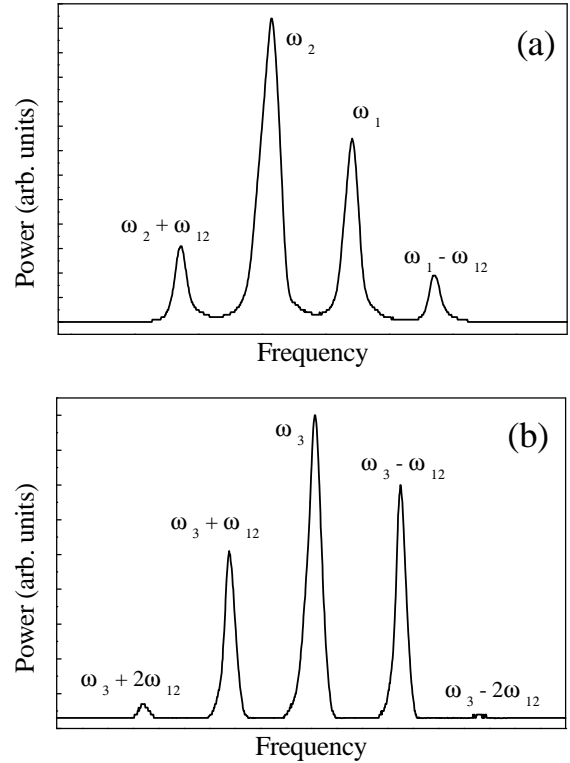,width=8cm}
\end{center}
\caption{\small{Typical oscilloscope signal showing the generation
of Raman sidebands for all three input frequencies. (a)
frequencies with $\sigma ^{+}$ polarization ($\omega _{1}$,
$\omega _{2}$ and their Raman sidebands) detected by SA1, (b)
frequencies with $\sigma ^{-}$ polarization ($\omega _{3}$ and its
Raman sidebands) detected by SA2.}} \label{fig3}
\end{figure}

Fig. 3 shows typical oscilloscope signals demonstrating the generation of
Raman sidebands for all three input frequencies. For the frequencies $\omega
_{1}$ and $\omega _{2}$ only the first-order sidebands (Stokes and
anti-Stokes fields, respectively) have been observed, while for the field $%
\omega _{3}$ up to the third-order sidebands have been seen, with the
higher-order sidebands appearing at larger input power. The intensity of the
generated sidebands grows almost linearly with the input power in the range
of up to 2 mW for each of the waves $\omega _{1}$, $\omega _{2}$ and $\omega
_{3}$. The minimum input power necessary for the generation of the
first-order sidebands was found in our experiment to be of the order of 10 $%
\mu W$ for each wave (the intensity is about 2 mW/cm$^{2}$). We stress that
the generation was achieved without the use of buildup cavities. We believe
that these results can be considered as an experimental confirmation of the
possibility for nonlinear-optical generation processes with a few photons.
With a pulse of a duration of a few $\mu s$ (as used, e.g., in experiment
Ref. \cite{hau99}) the intensity of 2 mW/cm$^{2}$ would correspond to the
energy of only a few light quanta per atomic cross-section. Thus, one is
approaching the regime of nonlinear quantum optics where a large
nonlinearity of the medium is created by single photons.

\begin{figure}[htb]
\begin{center}
\epsfig{figure=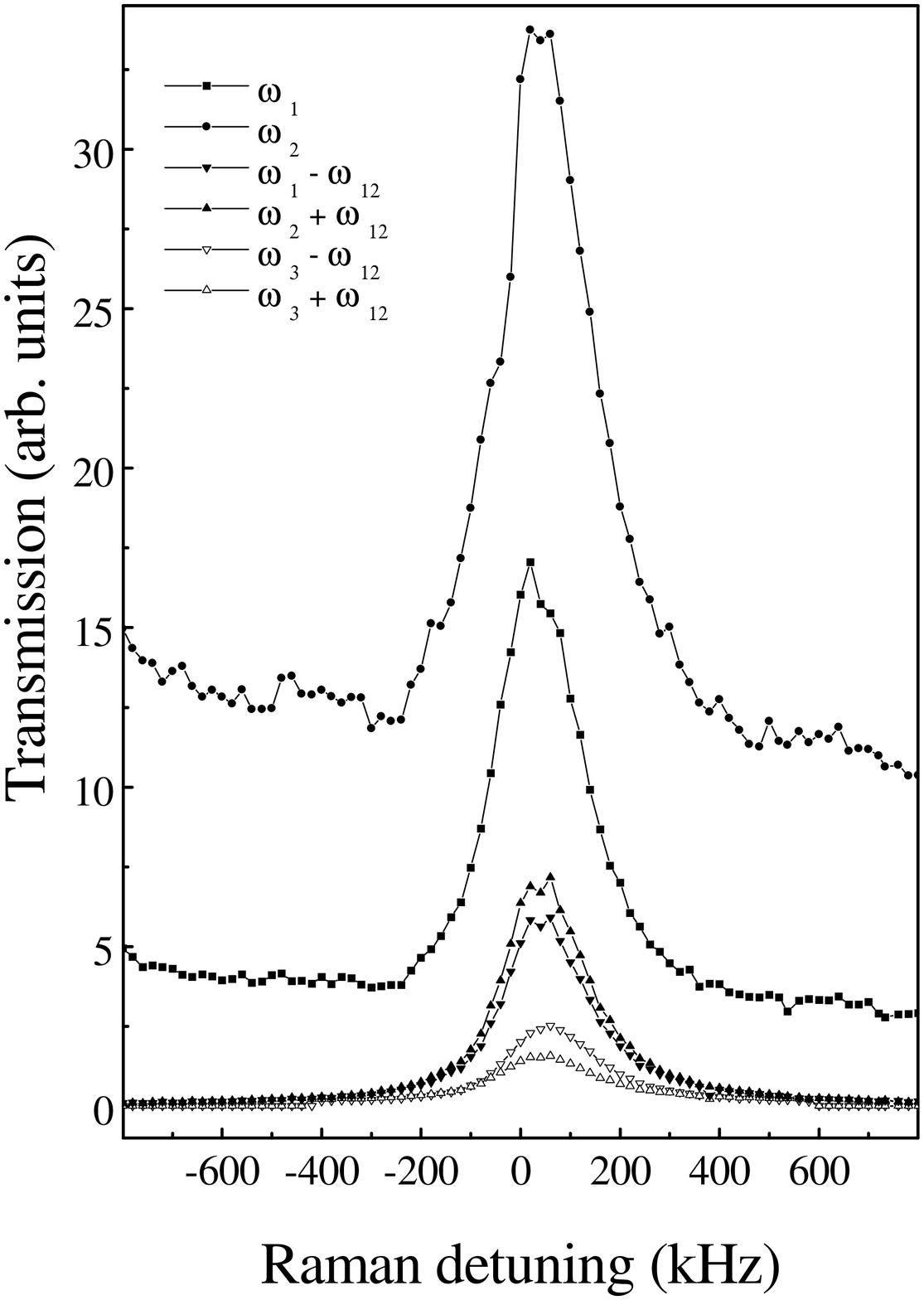,width=7.5cm}
\end{center}
\caption{\small{Dependence of the transmitted light power (plotted
on a linear scale) on the Raman detuning $\delta _{R}=(\omega
_{2}-\omega
_{1})-\omega _{12}$ ($\omega _{12}=1771.626$ MHz). Input power $%
P_{in}=700\mu W$ (intensity $I=150$ mW/cm$^{2}$, equal for each of
the input waves $\omega _{1}$, $\omega _{2}$ and $\omega _{3}$).
Temperature of the cell 190${{}^{\circ }}$C ($\tau =5.8$). The
frequency $\omega _{3}$ is
detuned by 3 GHz above the transition $3^{2}S_{1/2},F=2-3^{2}P_{1/2},F^{%
\prime }=2$ and passes the cell nearly unabsorbed.}} \label{fig4}
\end{figure}

The nonlinear generation at such low light intensities is already
an indirect confirmation of the EIT in action. For a direct
confirmation, we measured the dependence of all transmitted
frequencies on the AOM modulation frequency (Fig. 4). One can see
that the generation of all sideband frequencies occurs only in the
narrow range of Raman detuning $\delta _{R}=(\omega _{2}-\omega
_{1})-\omega _{12}$. This is exactly the same range where input
frequencies $\omega _{1}$ and $\omega _{2}$ experience reduced
absorption (EIT). The width of the range (10 - 200 kHz depending
on the input intensity) is much narrower than the natural width of
the excited state $3^{2}P_{1/2}$ (10 MHz), and its dependence on
the intensity of the
frequencies $\omega _{1}$ and $\omega _{2}$ follows the expected dependence $%
\delta _{EIT}=\Gamma +C\cdot I$. The generation peak is shifted from exact
Raman resonance due to both the buffer gas effects and the a.c. Stark shift
\cite{wyn97,wyn99}. The a.c. Stark shift is a differential shift of the
ground states $3^{2}S_{1/2},F=1$ and $F=2$ due to the off-resonance coupling
to the second excited state $3^{2}P_{1/2},F^{\prime }=1$ of the D$_{1}$ line
(which is 189 MHz apart from the $3^{2}P_{1/2},F^{\prime }=2$ state). Our
measurements show that this shift is proportional to the {\it input}
intensity of the frequencies $\omega _{1}$ and $\omega _{2}$ with a slope
depending on the optical density; for the parameters of Fig. 4 ($\tau
\approx 6$) the slope is about 0.2 kHz/(mW/cm$^{2}$). The total absorption
of $\omega _{1}$ and $\omega _{2}$ in Fig. 4 corresponds to 98\% outside the
EIT transparency window, while at resonance the absorption reduces only
moderately to 93\%. One must conclude therefore that it is not only
reduction of absorption, but also enhancement of the nonlinear
susceptibility that assist the generation.

\begin{figure}[htb]
\begin{center}
\epsfig{figure=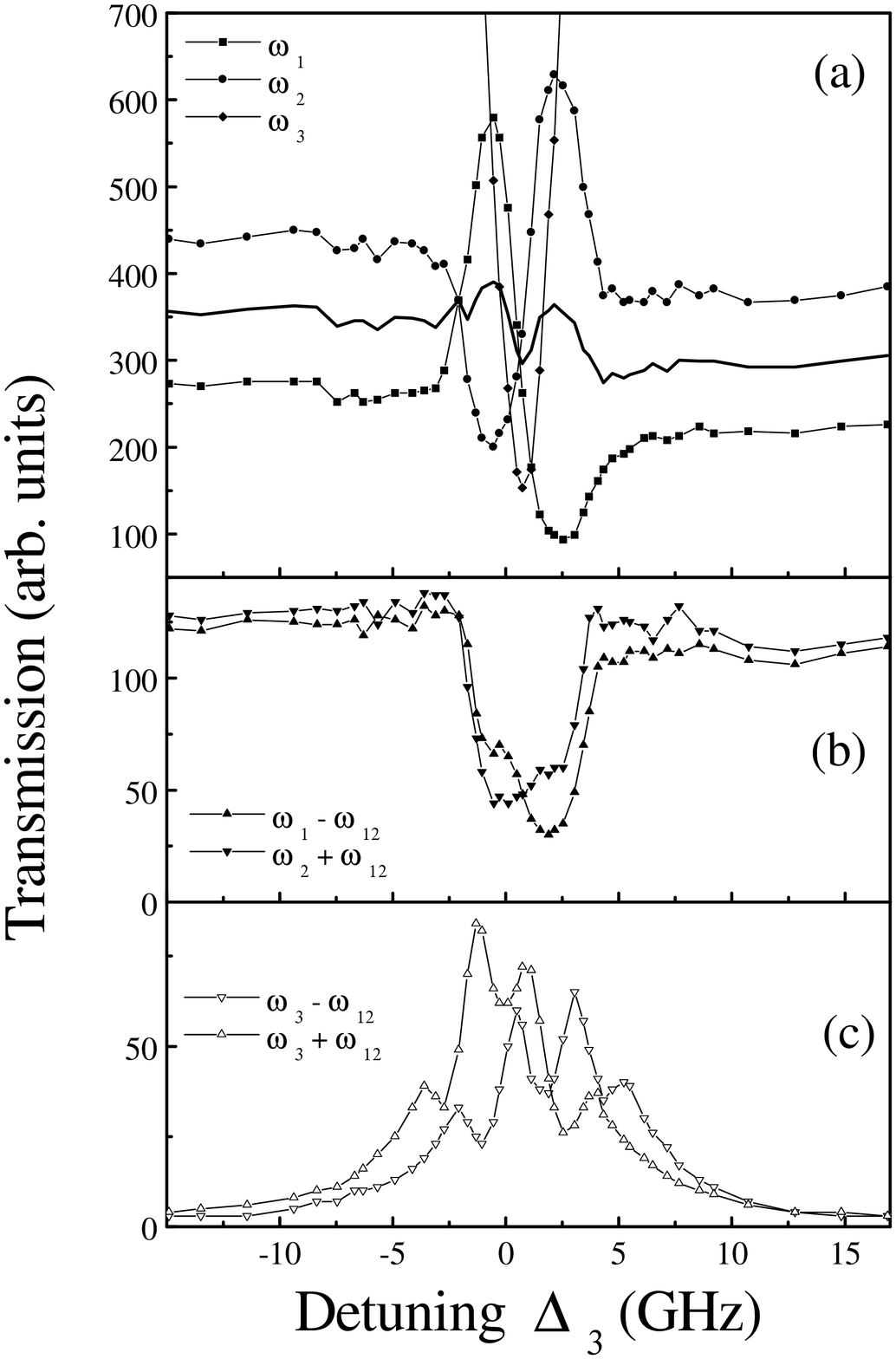,width=8cm}
\end{center}
\caption{\small{Dependence of the transmitted light power (plotted
on a linear scale) on the detuning $\Delta _{3}$ of frequency
$\omega _{3}$ from the transition
$3^{2}S_{1/2},F=2-3^{2}P_{1/2},F^{\prime }=2$. (a) For frequencies
$\omega _{1}$, $\omega _{2}$ and $\omega _{3}$. The solid curve is
the total transmitted intensity of $\omega _{1}$ and $\omega _{2}$
waves.
(b) For the first-order Raman sidebands of frequencies $\omega _{1}$ and $%
\omega _{2}$, (c) for the first-order Raman sidebands of frequency
$\omega
_{3}$. Input power $P_{in}=800\mu W$ (equal for each of the input waves $%
\omega _{1}$, $\omega _{2}$ and $\omega _{3}$). Temperature of the cell 190${%
{}^{\circ }}$C ($\tau =5.8$).}} \label{fig5}
\end{figure}

The intensity of both generated and transmitted pump fields considerably
depends on the value of frequency $\omega _{3}$. Figure 5 shows the
dependence of the intensities on detuning $\Delta _{3}$ of the $\omega _{3}$
wave from the transition $3^{2}S_{1/2},F=2-3^{2}P_{1/2},F^{\prime }=2$ in
the range of $\pm $16 GHz. The $\omega _{3}$ wave has almost no influence on
transmission of the resonant pump waves $\omega _{1}$ and $\omega _{2}$
(Fig. 5(a)) and their Raman sidebands (Fig. 5(b)) when $\left| \Delta
_{3}\right| $ is much larger than the Doppler width $\Delta _{D}\approx $1
GHz of the D$_{1}$ line. The $\omega _{3}$ wave itself is absorbed very
weakly in this range (therefore it is not shown in Fig. 5(a)). However, as
the resonance with the D$_{1}$ line is approached, the $\Delta _{3}$
dependence becomes more and more dramatic. Starting from the value $\left|
\Delta _{3}\right| $ of about 8 GHz, one can observe weak intensity
oscillations whose amplitude drastically increases for $\Delta _{3}$ being
in the immediate range of the resonance. When $\omega _{3}$ is tuned close
to $3^{2}S_{1/2},F=2-3^{2}P_{1/2}$ ($\Delta _{3}\approx -1.0\div 0$ GHz),
transmission of the $\omega _{2}$ wave decreases, while that of the $\omega
_{1}$ wave increases (Fig. 5(a)). This is because a larger part of atomic
population is pumped into the state $3^{2}S_{1/2},F=1$. Similar process
occurs at $\omega _{3}$ being tuned close to $3^{2}S_{1/2},F=1-3^{2}P_{1/2}$
($\Delta _{3}\approx 1.7\div 2.7$ GHz), where transmission of the $\omega
_{1}$ wave decreases, while that of the $\omega _{2}$ wave increases. It is
interesting, however, that at these detunings the total transmitted
intensity of the $\omega _{1}$ and $\omega _{2}$ waves (solid curve in Fig.
5(a)) slightly increases over the ''transparency level'' in the absence of
the $\omega _{3}$ wave. Despite of this fact, the intensity of generated
Stokes ($\omega _{1}-\omega _{12}$) and anti-Stokes ($\omega _{2}+\omega
_{12}$) fields diminishes sharply in the same range of detuning (Fig. 5(b)),
so that the total transmitted intensity of $\omega _{1}$, $\omega _{2}$ and
their sidebands remains approximately the same as for very large detunings.
This suggests that the absorption is not increased when the $\omega _{3}$
field is tuned on resonance, but the nonlinear susceptibility for the Stokes
and anti-Stokes fields decreases. At the same time, generation of Raman
sidebands of the $\omega _{3}$ field is maximized. All these facts indicate
that when $\omega _{3}$ is close to resonance with the D$_{1}$ line, the
processes of coherent scattering from each input wave into its Raman
sidebands become tightly coupled to one another and start to compete. Thus,
the decrease of the ($\omega _{1}-\omega _{12}$) and ($\omega _{2}+\omega
_{12}$) fields generation in the range of resonance may be explained by
direct competition of this process with the generation of ($\omega _{3}\pm
\omega _{12}$) fields. Inside the resonance range, the ($\omega _{1}-\omega
_{12}$) and ($\omega _{2}+\omega _{12}$) fields (Fig. 5(b)) reflect the
trend of the $\omega _{1}$ and $\omega _{2}$ fields (Fig. 5(a)),
respectively, i.e., increase of the $\omega _{1}$ intensity leads to
(moderate) increase of the ($\omega _{1}-\omega _{12}$) intensity, etc. On
the contrary, the behavior of the ($\omega _{3}\pm \omega _{12}$) sidebands
is more complicated: they reveal nice periodical oscillations with detuning $%
\Delta _{3}$, shifted in phase with respect to each other. These
oscillations are apparently related to those observed in \cite{mer99} and
also theoretically predicted in \cite{kor99} for the generation of ($\omega
_{3}+\omega _{12}$) frequency in double $\Lambda $ atoms, which show a
sinusoidal dependence of the generated wave intensity on detuning $\Delta
_{3}$ and optical density $\tau $. So, the shift of the ($\omega _{3}+\omega
_{12}$)-wave intensity with respect to the ($\omega _{3}-\omega _{12}$) one
may be explained simply by their different detunings from the resonance.

Finally, Fig. 6 demonstrates the measurement results for the optical density
dependence of all the waves, both pump and generated ones. The measurements
have been performed by taking the oscilloscope pictures at different cell
temperatures corresponding to different optical densities. The results
presented in Fig. 6 are obtained for the particular case when $\omega _{3}$
is tuned exactly on resonance with transition $%
3^{2}S_{1/2},F=2-3^{2}P_{1/2},F^{\prime }=1$. This configuration corresponds
to the all-resonant double $\Lambda $ system, where one expects generation
of the ($\omega _{4}=\omega _{3}+\omega _{12}$) wave resonant with the $%
F=1-F^{\prime }=1$ transition, and propagation dynamics leading to the
matching of the field Rabi frequencies to relation $g_{1}/g_{2}=g_{3}/g_{4}$
\cite{kor99,mer00}. However, we have not observed such a matching.
\begin{figure}[htb]
\begin{center}
\epsfig{figure=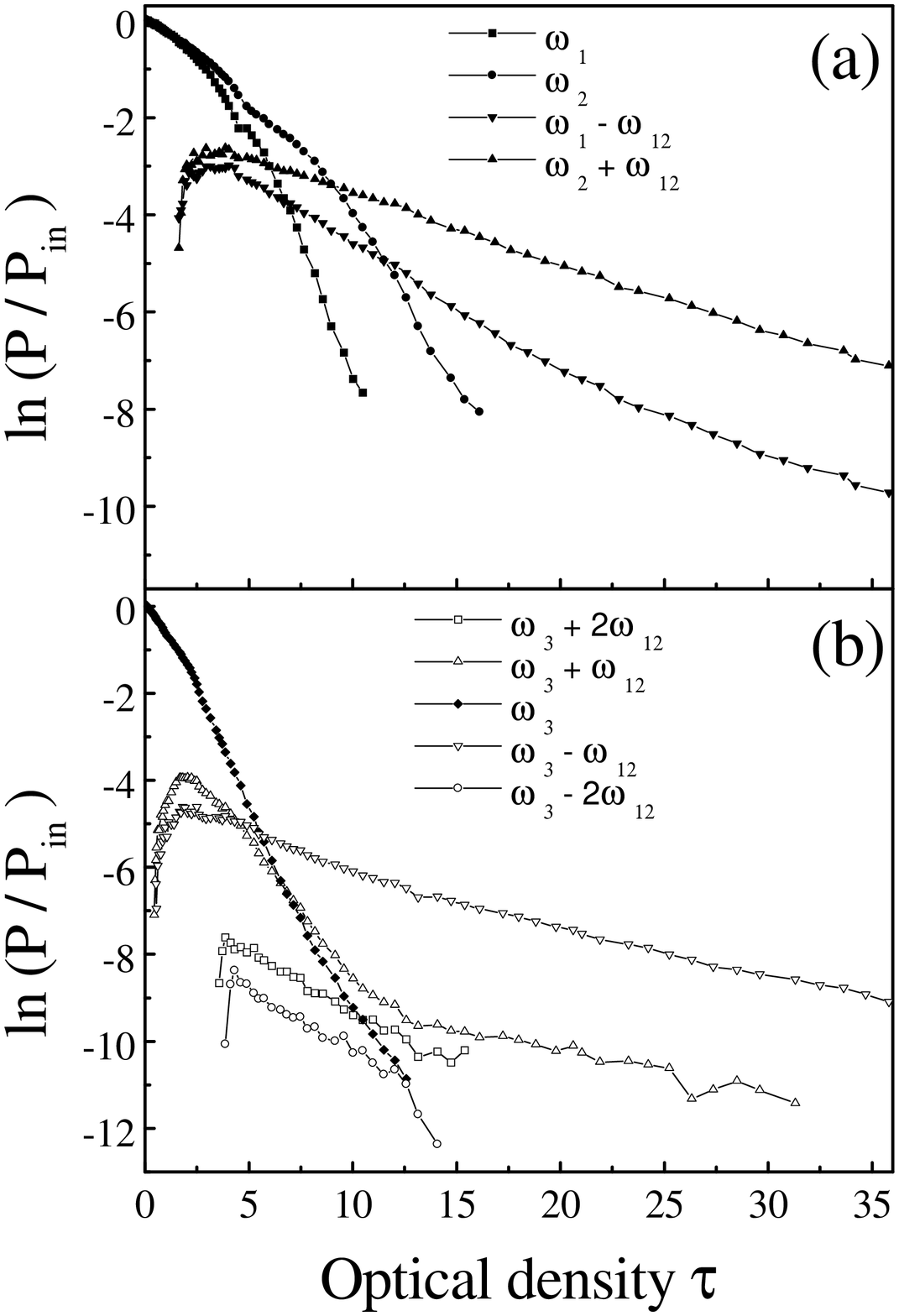,width=8cm}
\end{center}
\caption{\small{ Optical density dependence of the transmitted
light power normalized to the input power $P_{in}=450\mu W$ (equal
for each of the input
waves $\omega _{1}$, $\omega _{2}$ and $\omega _{3}$). (a) For frequencies $%
\omega _{1}$, $\omega _{2}$ and their Raman sidebands, (b) for frequency $%
\omega _{3}$ and its Raman sidebands. The frequency $\omega _{3}$
is tuned
exactly on resonance with the transition $3^{2}S_{1/2},F=2-3^{2}P_{1/2},F^{%
\prime }=1$.}} \label{fig6}
\end{figure}
One can
see from Fig. 6 that power of the pump waves $\omega _{1}$, $\omega _{2}$
and $\omega _{3}$ is attenuated monotonically with the optical density. This
attenuation occurs quite fast, with a rate being only slightly lower than
that given by Beer's exponential decay. There are at least two reasons for
this behavior. First of all, we don't have here real all-resonant double $%
\Lambda $ system since all the waves couple to all possible transitions
(which is, in fact, the reason for the broad Raman spectrum generation).
Therefore, the energy is transferred not only between the resonant fields,
but also goes to the off-resonant Raman sidebands. Second, the preparation
of the medium in the dark state relies in our case on optical pumping. In
this process, photons from the pump waves $\omega _{1}$, $\omega _{2}$ are
absorbed and then rescattered in part spontaneously to bring atomic
population in the dark state. The number of spontaneously scattered photons
is proportional to the excited state population which is in turn
proportional to the light intensity. Therefore, the pump beams experience
exponential losses during propagation in the medium. Nevertheless, the atoms
are prepared in the coherent superposition of both lower states $\left|
1\right\rangle $ and $\left| 2\right\rangle $, and as soon as they are
prepared, the generation of the Raman sidebands goes on very efficient. As
can be seen from Fig. 6, the first-order sidebands appear already at quite
small optical densities ($\tau =1\div 2$), grow very fast and reach their
maximum at densities of the order of $\tau =2.5\div 5$. The energy
conversion efficiency is approaching the value of 5-7\% for the sidebands of
$\omega _{1}$ and $\omega _{2}$ fields, and of 3-4\% for the first-order
sidebands of $\omega _{3}$ field. At this point of maximum generation, the
intensity of the first-order sidebands is large enough to induce the
generation of the second-order sidebands. Immediately after the initial,
very fast, stage of generation, the Raman sidebands start to decay. The
resonant sideband ($\omega _{3}+\omega _{12}$) is attenuated as fast as the
pump waves, while the other Raman sidebands decay slowly due to large
detuning from any resonance and, hence, small absorption. It is interesting
that different decay rates give rise to a curious situation at larger
optical densities when the generated sidebands are more intensive than the
pump fields.

\section{Conclusions}

In conclusion, we have experimentally demonstrated generation of Raman
sidebands in sodium atomic vapor excited on the D$_{1}$ line by resonant
c.w. optical fields of frequencies $\omega _{1}$ and $\omega _{2}$. The
first-order sidebands for frequencies $\omega _{1}$, $\omega _{2}$ and up to
the third-order sidebands for the probe field of frequency $\omega _{3}$
have been observed. The efficient generation takes place due to the
preparation of atoms in a dark superposition leading to reduced absorption
and enhanced nonlinear susceptibility. This has been directly confirmed by
measuring the frequency difference ($\omega _{1}-\omega _{2}$) dependence of
the transmission, which evidences that the generation of all sidebands as
well as the reduced absorption of pump fields occur only in the narrow range
around Raman resonance $\omega _{1}-\omega _{2}=\omega _{12}$. Since the
decay rate of the dark state is only a few kHz in our experiment, the
generation is efficient even at very low pump powers, with the threshold
value being of 10 $\mu W$ for each input wave. Our measurements show that
the Raman sidebands are generated at any value of the frequency $\omega _{3}$%
. However, the generation of Raman sidebands of the $\omega _{3}$ field is
maximized and competes with generation of the sidebands of $\omega _{1}$ and
$\omega _{2}$ fields when $\omega _{3}$ is close to resonance with the D$_{1}
$ line. Inside this resonance range, the sideband intensities reveal
periodical oscillations with $\omega _{3}$. In the present scheme, the
coherent medium is prepared with the c.w. radiation by means of the
dissipative process of optical pumping. Therefore, a large part of the
radiation energy is lost. This is reflected in the optical density
dependence exhibiting fairly fast and almost exponential decay of the
resonant pump waves. We believe that much smaller energy losses and,
correspondingly, higher conversion efficiency may be achieved by use of the
adiabatic passage technique for the preparation of phaseonium \cite{apt}.
Together with a further reduction of the dark state decay rate (which is
certainly possible by using better magnetic field shielding, stabilizing AOM
frequency and optimizing the buffer gas pressure), this should allow to
readily approach the regime of a few-photon nonlinear optics. Our
experimental results also give rise to the challenge of developing a theory
of Raman sideband generation by resonant c.w. radiation that will provide an
understanding of a complicated interplay of the participating e.m. waves.

This work was supported by the Austrian Science Foundation under project No.
P 12894.

\end{document}